\documentclass[10pt,journal,compsoc]{IEEEtran}

\ifCLASSOPTIONcompsoc
  \usepackage[nocompress]{cite}
\else
  \usepackage{cite}
\fi

\usepackage[utf8]{inputenc}
\usepackage[english]{babel}
\usepackage{amsmath}

\usepackage{graphicx}
\usepackage{caption}
\usepackage{listings}
\usepackage{float}
\usepackage{algorithm2e}
\usepackage{caption}
\usepackage{array}
\usepackage{booktabs}

\usepackage{hyperref}
\usepackage{makecell}

\usepackage[table]{xcolor}

\usepackage{cardEstMacrosTikz}
\usepackage{cardEstMacros}

\usepackage{lettrine}
\usepackage{subfigure}
\usepackage{multirow}
\usepackage{multicol}
\usepackage{tablefootnote}
\usepackage{enumerate}
\usepackage{enumitem}
\usepackage{extarrows}

\usepackage{booktabs}
\usepackage{mathtools}
\usepackage{multirow}
\usepackage{subfigure}
\usepackage{caption}
\usepackage{graphicx}
\usepackage{url} 
\usepackage{pifont}
\usepackage{extarrows}
\usepackage{amsfonts}
\usepackage{marvosym}

\begin{document}
\title{Cross-domain Augmentation Networks for Click-Through Rate Prediction}
\author{Xu Chen, Zida Cheng, Shuai Xiao\textsuperscript{\Letter}, Xiaoyi Zeng, Weilin Huang
\IEEEcompsocitemizethanks{\IEEEcompsocthanksitem Xu Chen, Zida Cheng, Shuai Xiao, Xiaoyi Zeng and Weilin Huang are with Alibaba Group. E-mail: \{huaisong.cx, chengzida.czd, shuai.xsh, weilin.hwl\}@alibaba-inc.com, yuanhan@taobao.com. \textsuperscript{\Letter} indicates the corresponding author}
\thanks{}
}

\markboth{}%
{}
%

\IEEEtitleabstractindextext{%
\begin{abstract}
Data sparsity is an important issue for click-through rate (CTR) prediction, particularly when user-item interactions is too sparse to learn a reliable model. Recently, many works on cross-domain CTR (CDCTR) prediction have been developed in an effort to leverage meaningful data from a related domain. 
However, most existing CDCTR works have an impractical limitation that requires homogeneous inputs (\textit{i.e.} shared feature fields) across domains, and CDCTR with heterogeneous inputs (\textit{i.e.} varying feature fields) across domains has not been widely explored but is an urgent and important research problem.
In this work, we propose a cross-domain augmentation network (CDAnet) being able to perform knowledge transfer between two domains with \textit{heterogeneous inputs}. 
Specifically, CDAnet contains a designed translation network and an augmentation network which are trained sequentially.
The translation network is able to compute features from two domains with heterogeneous inputs separately by designing two independent branches, 
and then learn meaningful cross-domain knowledge using a designed cross-supervised feature translator. Later the augmentation network encodes the learned cross-domain knowledge via feature translation performed in the latent space and fine-tune the model for final CTR prediction.
Through extensive experiments on two public benchmarks and one industrial production dataset, we show CDAnet can learn meaningful translated features and largely improve the performance of CTR prediction. CDAnet has been conducted online A/B test in image2product retrieval at Taobao app over 20 days, bringing an absolute \textbf{0.11 point} CTR improvement and a relative \textbf{1.26\%} GMV increase.
\end{abstract}

}

\maketitle

\IEEEdisplaynontitleabstractindextext

%
\IEEEpeerreviewmaketitle

\section{Introduction}
\lettrine[lines=2]{C}{lick-through} rate (CTR) prediction which estimates the probability of a user clicking on a candidate item, has a vital role in online services like recommendation, retrieval, and advertising~\cite{cheng2016wide,zhou2018deep,ouyang2019deep}. 
For example, Taobao, as one of the largest e-commercial platforms in the world, has a list of application domains such as text2product retrieval and image2product retrieval. Each domain has its own CTR prediction model, which is termed as single-domain CTR prediction. 
It has been pointed out that data sparsity is a key issue that significantly limits the improvement of single-domain CTR models~\cite{li2015click}, and recent effort has been devoted to improving the CTR models by leveraging data from the other related domains~\cite{li2015click,ouyang2020minet,liu2022continual,zhang2022keep,sheng2021one}. 
\begin{figure}[t]
\centering
\begin{minipage}[t]{0.24\textwidth}
\centering
\includegraphics[width=0.85\textwidth]{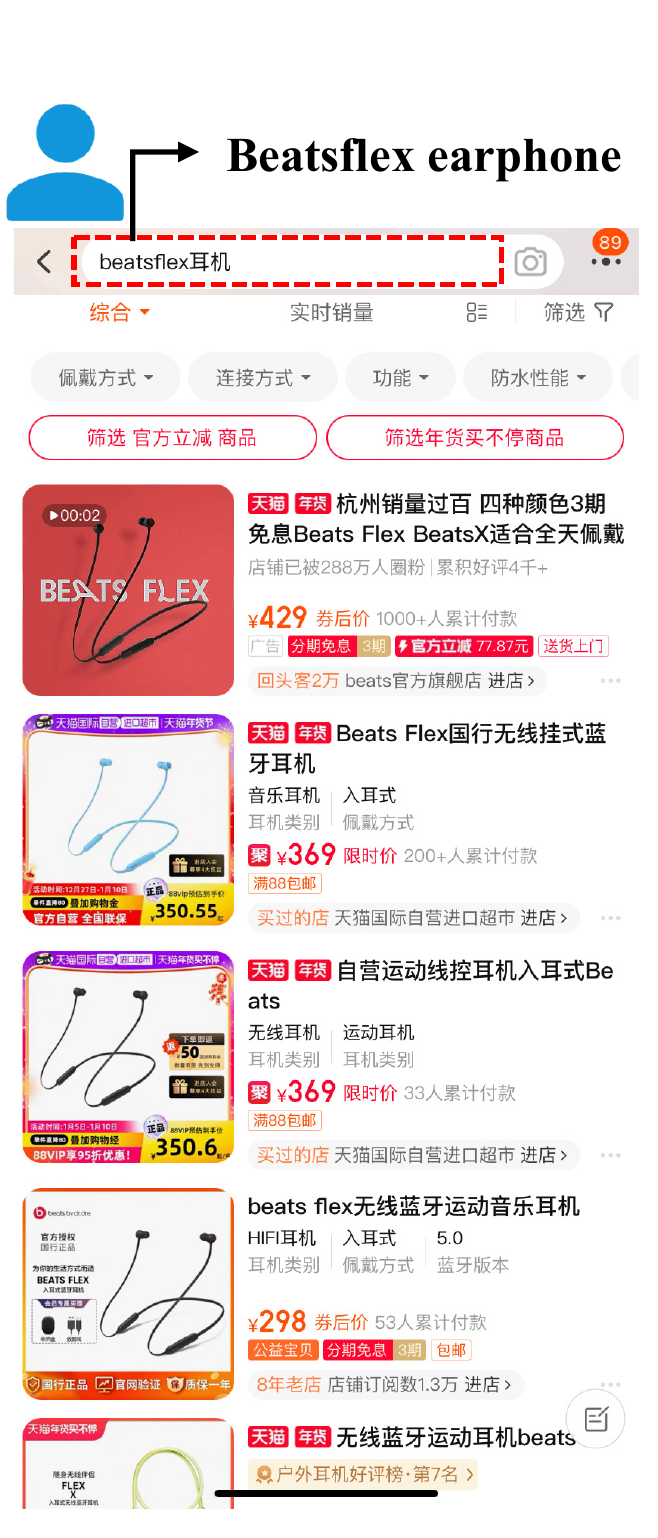}
\caption*{(a) text2product retrieval}
\end{minipage}
\begin{minipage}[t]{0.24\textwidth}
\centering
\includegraphics[width=0.85\textwidth]{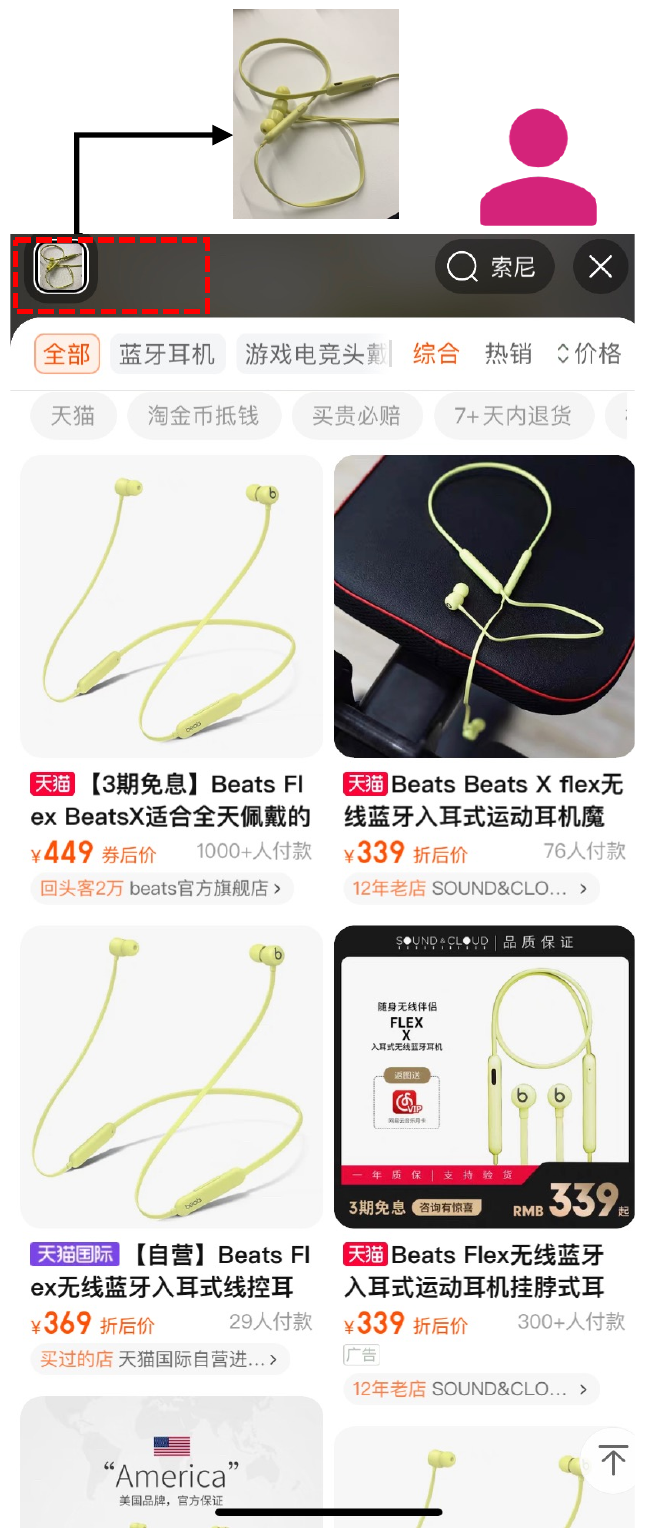}
\caption*{(b) image2product retrieval}
\end{minipage}
\caption{An example of text2product retrieval and image2product retrieval at Taobao app.
\label{figure:taobao_app_example}
}
\end{figure}


Various cross-domain CTR prediction (CDCTR) methods have recently been developed, which can be roughly categorized into two groups: joint training and pre-training \& fine-tuning. 
%
The joint training approach is developed by combining multiple CTR objectives from different domains into a single optimization process. It usually has shared network parameters to build connections, and transfer the learned knowledge across different domains like MiNet~\cite{ouyang2020minet}, DDTCDR~\cite{li2019ddtcdr} and STAR~\cite{sheng2021one}. However, since two domains usually have different objectives in optimization, these methods usually suffer from an optimization conflict problem, which might lead to a negative transfer result~\cite{sener2018multi,vandenhende2021multi}. To deal with this issue, a number of recent approaches have been proposed for jointly training multiple CTR objectives~\cite{ma2018modeling,tang2020progressive,lin2019pareto}. 
On the other hand, the pre-training \& fine-tuning methods often have two stages, by training a CTR model sequentially in a source domain and then in a target domain, where the performance in the target domain can be generally improved by leveraging model parameters pre-trained from the source domain. 
Notice that only one objective is utilized for optimization in each training stage, which significantly reduces the impact of negative transfer. Therefore, this method has been widely applied at Taobao platform~\cite{liu2022continual,zhang2022keep}.

Importantly, most of recent CDCTR methods have been developed to explore additional cross-domain data with \textit{homogeneous input features}~\cite{ma2018modeling,tang2020progressive,ouyang2020minet,sheng2021one,liu2022continual,zhang2022keep}, which means the feature fields across domains are shared. For example, both MiNet~\cite{ouyang2020minet} and MMOE~\cite{ma2018modeling} have a shared embedding layer of inputs to transfer knowledge across domains. Particularly, recent pre-training \& fine-tuning methods presented in~\cite{liu2022continual,zhang2022keep} also require homogeneous inputs that enables them to apply the data from a target domain directly to the pre-trained source model, and thus is able to generate additional features for fine-tuning the target model.   
However, the requirement with homogeneous inputs might largely limit their applications in practice. Many cross-domain scenarios often exist \textit{heterogeneous input features} which means two domains have varying feature fields.
For instance, for image2product retrieval in Taobao, image is an important query information, while text plays as the key query information in text2product retrieval.
An image retrieval system relies on image features a lot, while a text retrieval system focuses on text features heavily.
%
At Taobao app, text2product retrieval has developed for many years and generated an order of magnitude more data than that of image2product retrieval. It is desirable to perform knowledge transfer that learns user behaviors from text2product domain, and then apply them for improving the performance on image2product retrieval.
%

This inspired us to develop a new CDCTR approach allowing for \textit{heterogeneous input features}, which has not been widely studied yet but is challenging and urgent to be solved in industrial applications. A straightforward solution is to build on recent works~\cite{li2019ddtcdr,ma2018modeling,tang2020progressive}, by replacing the shared-embedding layers with multiple domain-specific layers which process heterogeneous inputs separately, but it would inevitably break the key module designed specifically for transferring knowledge, resulting in unsatisfied performance. The up-to-date STAR~\cite{sheng2021one} can work with heterogeneous inputs, but it mainly relies on parameter-sharing to transfer knowledge and simultaneously involves multiple CTR objectives in learning, which may cause ineffective knowledge transfer. 
%
%


\begin{figure*}[t]
\centering
\begin{minipage}[t]{0.22\textwidth}
\centering
\includegraphics[width=\textwidth]{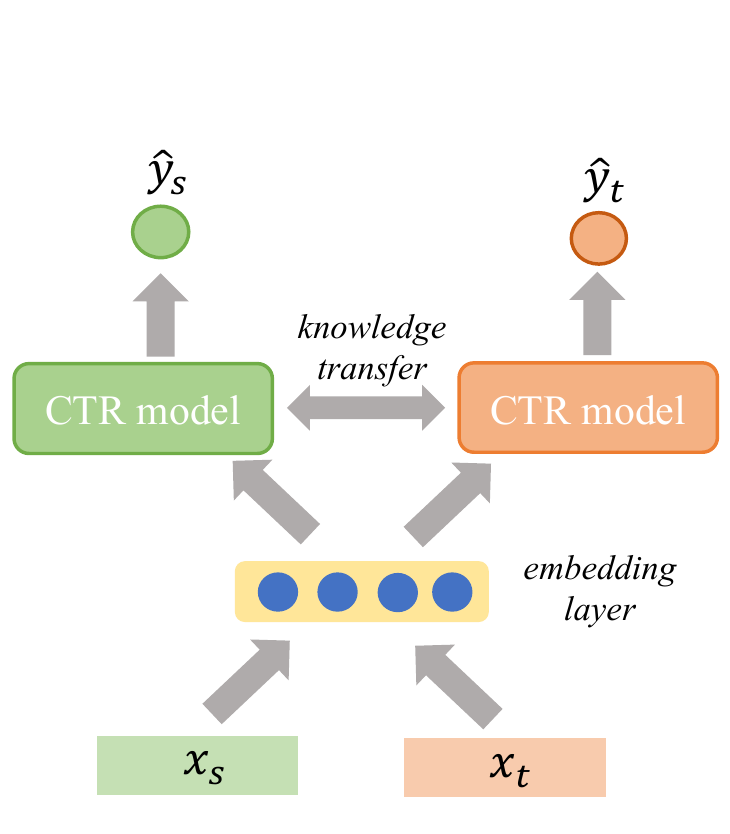}
\caption*{(a) Joint training}
\end{minipage}
\hspace{5pt}
\begin{minipage}[t]{0.24\textwidth}
\centering
\includegraphics[width=\textwidth]{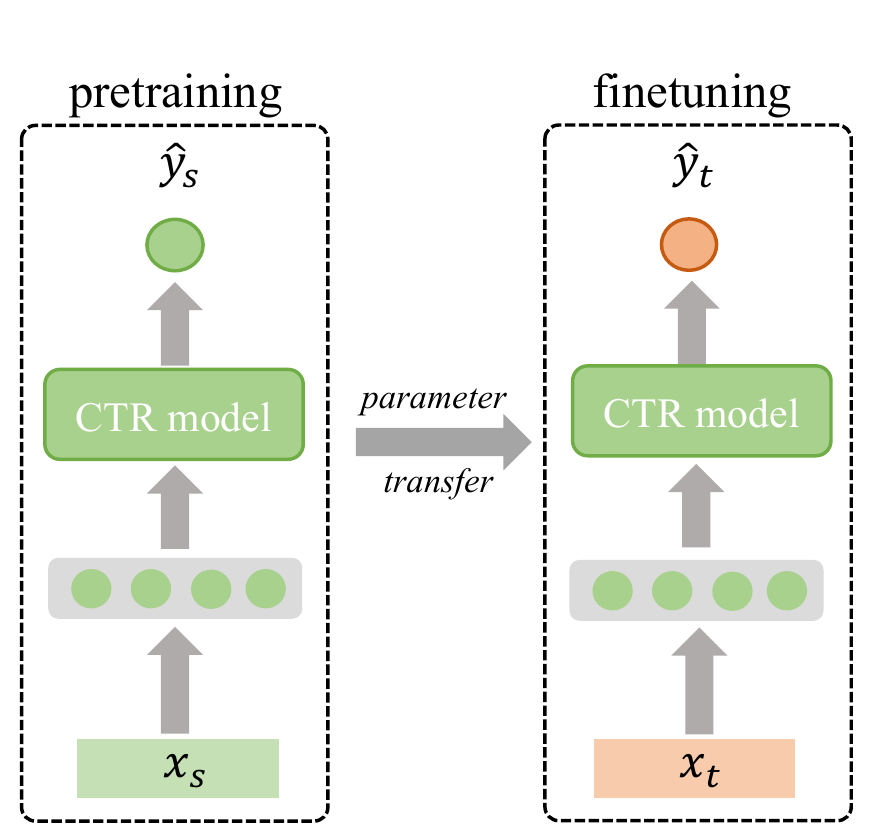}
\caption*{(b) pre-training \& fine-tuning}
\end{minipage}
\begin{minipage}[t]{0.42\textwidth}
\centering
\includegraphics[width=\textwidth]{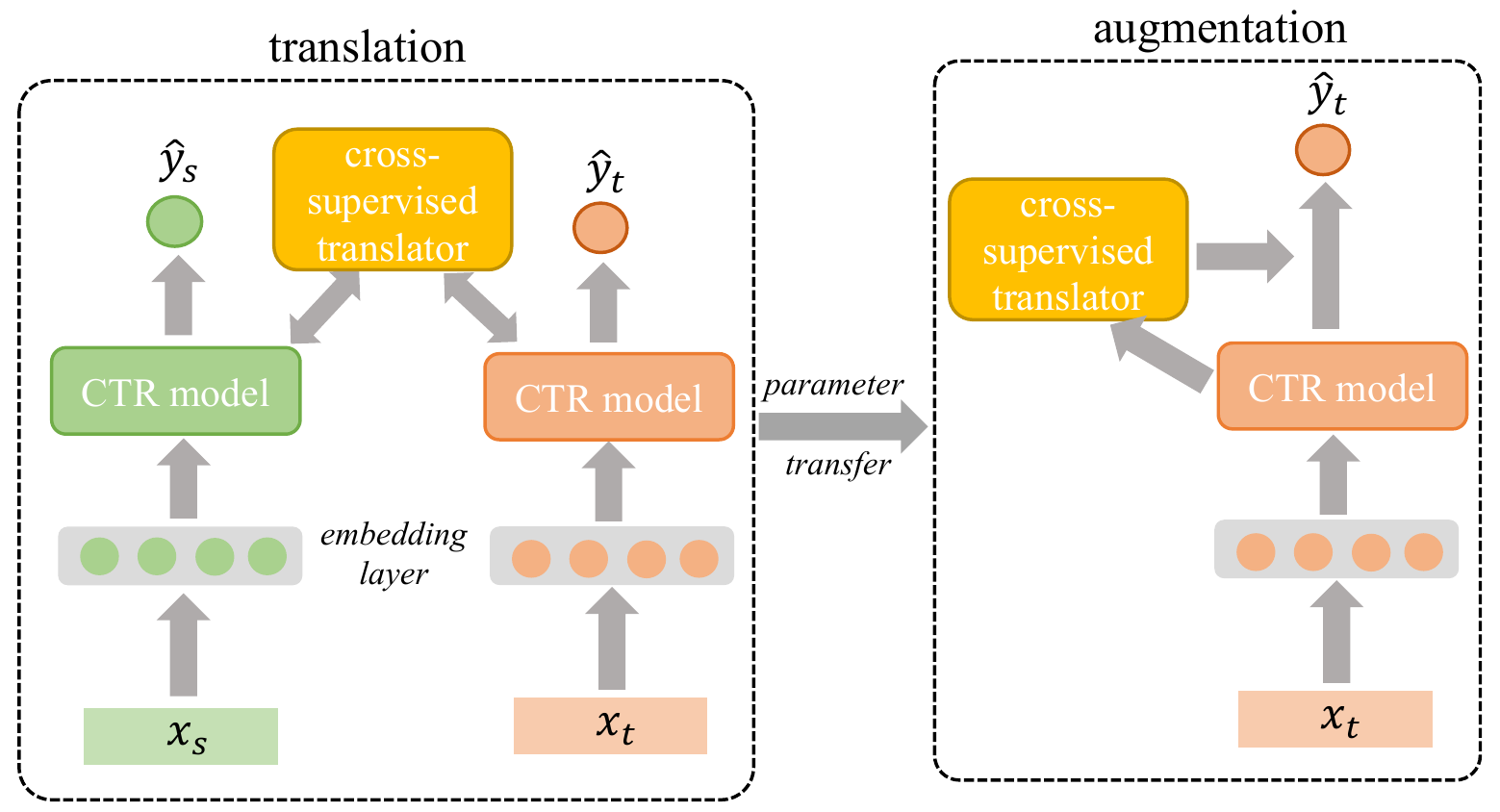}
\caption*{(c) CDAnet}
\end{minipage}
\caption{The comparison of joint training, pre-training \& fine-tuning and our CDAnet. a) is the joint learning scheme where the embedding layer is usually shared and knowledge transfer techniques are employed between two CTR domains. b) shows the pre-training \& fine-tuning learning style where a source domain CTR model will first be trained. Then the target domain model will load the pre-trained source domain model parameters and fine-tune itself with the target domain objective. c) shows the learning style of CDAnet. First, the translation network learns how to map the latent features from target domain to source domain by a cross-supervised translator. Then the augmentation network reuses the pre-trained parameters and employs the translated latent features as additional information to perform cross-domain augmentation for target domain CTR prediction.
}
\label{figure:general_comparison}
\end{figure*}
In this paper, we propose novel cross-domain augmentation networks (CDAnet) consisting of a translation network and an augmentation network, which are performed sequentially by first learning cross-domain knowledge and then implicitly encoding the learned knowledge into the target model via cross-domain augmentation. 
Specifically, a translation network is designed to process the inputs from different domains separately (which allows for heterogeneous input features), and knowledge translation is learned in the latent feature space via a designed cross-supervised feature translator. 
Then cross-domain augmentation is performed in the augmentation network by augmenting target domain samples in latent space with their additional translated latent features.
This implicitly encodes the knowledge learned from the source domain, providing diverse yet meaningful additional information for improving the fine-tuning on the target model. 
%
A learning comparison between CDAnet and other methods is shown in Figure~\ref{figure:general_comparison}.
Through experiments, we demonstrate that CDAnet can largely improve the performance of CTR prediction, and it has been conducted online A/B test in image2product retrieval at Taobao app, bringing an absolute \textbf{0.11 point} CTR improvement, with a relative \textbf{1.26\%} GMV increase. It has been successfully deployed online, serving billions of consumers.
In a nutshell, the contributions of this work are summarized as follows:
\begin{itemize}
  \item We identify a new cross-domain CTR prediction with \textit{heterogeneous inputs}, which allows the target model to learn meaningful additional knowledge from a different domain. This addresses the issue of data sparsity efficiently on CTR prediction, and also set it apart from most existing cross-domain methods only allowing for homogeneous inputs.
    \item We propose cross-domain augmentation networks (CDAnet) consisting of a designed translation network and an augmentation network. The translation network is able to learn cross-domain knowledge from  heterogeneous inputs by computing the features of two domain separately. Then it works as an efficient domain translator that encodes the learned knowledge implicitly in the latent space via feature augmentation, during the fine-tuning of target model.
    %
    %
    %
    \item Extensive experiments on various datasets demonstrate the effectiveness of the proposed method. Our CDAnet has been deployed in image2product retrieval at Taobao app, and achieved obvious improvments on CTR and GMV. In addition, through empirical studies, we show that CDAnet can learn meaningful translated features for boosted CTR improvement.
\end{itemize}

 

\section{Related Work}
\subsection{CTR Prediction}
\textbf{Single-domain CTR:} Click-through rate (CTR) prediction plays a vital role in various online services, such as modern search engines, recommendation systems and online advertising. Previous works usually combine logistic regression~\cite{richardson2007predicting} and feature engineering for CTR prediction. These methods often lack the ability to model sophisticated feature interactions, and heavily rely on human labor of designing features.
With the excellent feature learning ability of deep neural networks (DNN), deep learning approaches have been extensively studied on CTR prediction, and recent works focus on applying DNN for learning feature interactions, such as Wide\&Deep~\cite{cheng2016wide}, DeepFM~\cite{10.5555/3172077.3172127} and DCN~\cite{wang2017deep}.
For example, in~\cite{cheng2016wide}, Cheng \textit{et al.} combined  shallow linear models and deep non-linear networks to capture both low and high-order features, while the power of factorization machine~\cite{rendle2010factorization} and deep networks are combined for CTR prediction in~\cite{10.5555/3172077.3172127}. 
In DCN~\cite{wang2017deep}, Wang \textit{et al.} designed a deep \& cross network to learn bounded-degree feature interactions. 
Furthermore, deep models also demonstrate a strong capability for modeling richer information for CTR tasks.
For example, DIN~\cite{zhou2018deep} and DIEN~\cite{zhou2019deep} were proposed to capture user interests based on historical click behaviors, and DSTN~\cite{ouyang2019deep} takes the contextual ads when modeling user behaviors.

\textbf{Cross-domain CTR:} Single-domain CTR prediction suffers from a data sparsity issue because user behaviors in a real-world system are usually extremely sparse. Accordingly, cross-domain CTR prediction is developed to leverage user behaviors in a relevant but data-rich domain to facilitate learning in the current domain.
A joint learning method was recently developed and has become a representative approach for cross-domain CTR. For example, STAR~\cite{sheng2021one} is a star topology model that contains a centered network shared by different domains, with multiple domain-specific networks tailored for each domain. 
In MiNet~\cite{ouyang2020minet}, Ouyang \textit{et al.} attempted to explore auxiliary data (e.g. historical user behaviors and ad title) from a source domain to improve the performance of a target domain. 
Meanwhile, a number of cross-domain recommendation (CDR) methods have been developed~\cite{hu2018conet,li2019ddtcdr,yuan2019darec,chen2020towards}, which can be naturally introduced to cross-domain CTR problems. For instance, a deep cross connection network is introduced in~\cite{hu2018conet}, and it is able to transfer user rating patterns across different domains. In DDTCDR~\cite{li2019ddtcdr}, Li \textit{et al.} proposed a deep dual transfer network that can bi-directionally transfer information across domains in an iterative style. However, such models have two different CTR objectives during joint learning, which might lead to the problem of gradient interference~\cite{sener2018multi,wang2020gradient,yu2020gradient}. 
To handle the limitations, 
MMOE~\cite{ma2018modeling} was developed to learn an adaptive feature selection for different tasks, by using a shared mixture-of-expert model with task-specific gating networks. It explicitly models task relationships dynamically, allowing for automatically allocating model parameters which alleviates task conflicts in optimization.
To decouple learning task-specific and task-shared information more explicitly, PLE~\cite{tang2020progressive} separates the network of task-shared components and task-specific components, and then adopts a progressive routing mechanism able to extract and separate deeper semantic knowledge gradually. 

Apart from joint training, pre-training \& fine-tuning is a widely-applied two-stage learning paradigm. In the pre-training stage, a model is first trained in a source domain. Then in fine-tuning stage, a target model would load the pre-trained model parameters, and then fine-tune itself for target domain CTR prediction. In each stage, only one objective is used for optimization, and thus the gradient interference issue can be alleviated to some extent. It has been shown that this method can be more efficient and effective for industrial systems~\cite{chen2021user}. 
Recently, Zhang \textit{et al.} proposed KEEP ~\cite{zhang2022keep} which is a two-stage framework that consists of a supervised pre-training knowledge extraction module performing on web-scale and long-time data, and a plug-in network that incorporates the extracted knowledge into the downstream fine-tuning model. 
In CTNet~\cite{liu2022continual}, Liu \textit{et al.} focus on the CDCTR problem in a time-evolving scenario.
In addition, a number of recent works such as~\cite{evci2022head2toe,sung2022lst} investigate diffident layers of the network with various transfer manners (e.g. linear probing and fine-tuning), to perform knowledge transfer more efficiently.

Most of these CDCTR works focus on transferring knowledge between different user behavior distributions while ignoring the difference of input features, so they usually have a clear limitation on cross-domain homogeneous inputs~\cite{ma2018modeling,tang2020progressive,liu2022continual,zhang2022keep}. By contrast, the proposed CDAnet considers both the difference of user behavior distributions and input features in learning. It allows heterogeneous inputs and designs novel translation and augmentation networks to perform more effective knowledge transfer for this scenario.



\subsection{Translation Learning}
Translation is a task that translates the data from a source domain to a target domain while preserving the content information, which has two hot research topics: image2image translation~\cite{pang2021image} and neural machine translation~\cite{otter2020survey}. As neural machine translation works on sequential data and is not quite related to our work, we will mainly introduce image2image translation here.
In earlier works of image2image translation, researchers usually use aligned image pairs to train the translation model. Pix2pix~\cite{isola2017image} is the first unified image2image translation framework based on conditional GANs~\cite{mirza2014conditional}. However, the paired images are hard to be obtained and then the unpaired image2image translation networks are proposed~\cite{liu2017unsupervised,yi2017dualgan,zhu2017unpaired,kim2017learning}. In unpaired image2image translation, some works propose the constraint of preserving the image properties of the source domain data such as semantic features~\cite{taigman2016unsupervised} and class labels~\cite{bousmalis2017unsupervised}. Another well-known and effective constraint is the cycle-consistency loss~\cite{yi2017dualgan,zhu2017unpaired,kim2017learning} in which if an input image is translated to the target domain and back, we can obtain the original input image. 
Despite the various works on image2image translation, an indispensable part among them is the cross translation part which translates the image of the source domain to the image in the target domain. Our CDAnet also has a similar cross-supervised translator. Whereas, the key difference is CDAnet does not aim to obtain the translated result in data space but to learn the mapped latent features to augment the CTR prediction in the target domain.

\begin{figure*}[t]
\centering
\includegraphics[width=13.0cm]{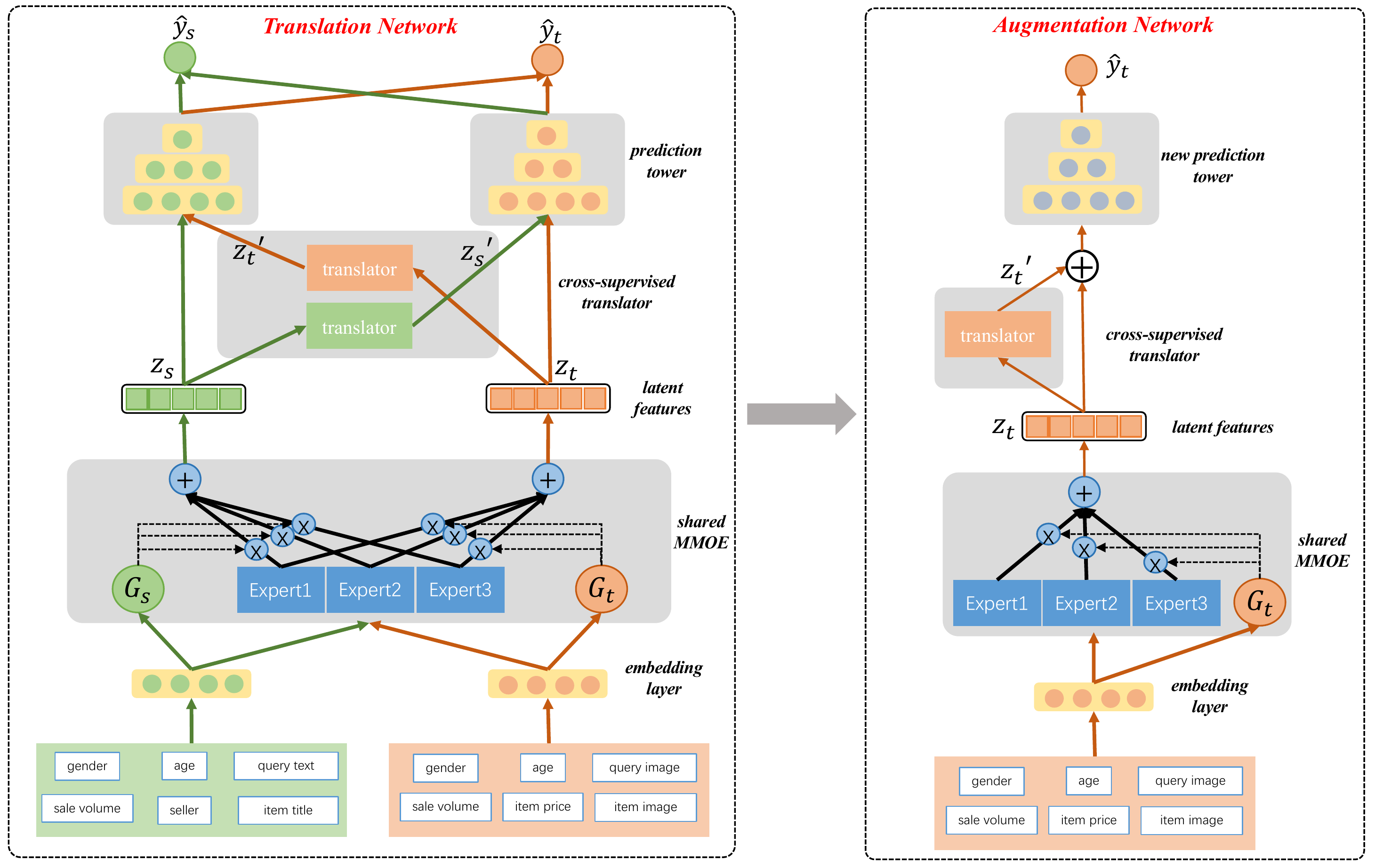}
\caption{The architecture of our cross-domain augmentation networks (CDAnet). First, the translation network mainly focuses on learning the translator by cross supervision. Then the translation network parameters except the tower layer are transferred to the augmentation network. After this, the augmentation network takes the target domain samples and their translated latent features together to augment the fine-tuning of the target domain model.}
\label{figure:model_architecture}
\end{figure*}
\section{Problem Formulation}
In CDCTR prediction with heterogeneous input features, given source domain $\mathcal{S}$, we have its training samples $(x_{s},y_{s})$ where $x_{s} \in \mathbb{R}^{F_{s}\times 1}$ denotes the input features and $y_{s}\in \{0,1\}$ is the click label (i.e. 1 means click while 0 means non-click). 
Similarly, we have the target domain $\mathcal{T}$ and its training samples $(x_{t},y_{t})$ in which $y_{t}\in \{0,1\}$ is the click label in target domain and $x_{t}\in \mathbb{R}^{F_{t}\times 1}$ is the input features. $F_{s}$ and $F_{t}$ denote the input feature dimension of source and target domain, respectively.
We define $x_{s}$ and $x_{t}$ are heterogeneous when they have varying feature fields. For instance, the text query is important and widely used in text2product retrieval. While in image2product retrieval, we do not have the text query but have image query and the image feature is quite different from the text. When two domains have heterogeneous input features, many recent works cannot well handle this and it is challenging to bridge the heterogeneous gap for efficient knowledge transfer. 

\section{Method}
\subsection{Overview}
The key of modeling in CDCTR with heterogeneous input features is to make sure that the model can embed the heterogeneous inputs and meanwhile have a good ability of transferring knowledge even without the widely used shared embedding layer technique appeared in recent works~\cite{ouyang2020minet,liu2022continual,ma2018modeling}. Accordingly, we propose our CDAnet which consists of two sequentially learned networks: translation network and augmentation network.
First, the translation network embeds the heterogeneous input features with decoupled embedding layers and learns how the latent features are translated mutually by a designed cross-supervised translator. Then, the pre-trained parameters of translation network are transferred to the augmentation network. Next, the augmentation network will combine the translated latent features and the original latent features of target domain samples together to fine-tune the target domain model. The model architecture is shown in Figure~\ref{figure:model_architecture}. Details about each module are demonstrated in the following parts.

\subsection{Translation Network}
Unlike image2image translation which aims to translate images in data space, our translation network learns to translate the latent features between domains. It has four parts including decoupled embedding layer, MMOE-based feature extractor, cross-supervised translator and prediction tower.

\subsubsection{Decoupled Embedding Layer}
Since the input features are heterogeneous, it is necessary to decouple the embedding layer of two domains so that the embedding layer can adapt to the input patterns in its own domain. Given the inputs $x_{s}\in \mathbb{R}^{F_{s}\times 1}$ of source domain and $x_{t}\in \mathbb{R}^{F_{t}\times 1}$ of the target domain, the embedding layer $H_{s}$ and $H_{t}$, we have:
\begin{align}
    h_{s}=H_{s}(x_{s}),~h_{t}=H_{t}(x_{t})
\end{align}
where $h_{s}\in \mathbb{R}^{d\times 1}$ and $h_{t}\in \mathbb{R}^{d\times 1}$ are the embedding features of $x_{s}$ and $x_{t}$, respectively. $d$ is the latent feature dimension. $H_{s}$ and $H_{t}$ are domain-specific embedding layers.

\subsubsection{MMOE-based Feature Extractor}
In translation network, the two CTR objectives of source domain and target domain are jointly optimized. They have different optimization directions because of the different data distributions. Ignoring this may cause the optimization conflict problem and lead to negative transfer. Here, we place a shared MMOE-based network to extract useful features for each domain so that each domain can automatically have its own parameters. We denote $F_{i}$ as the $i$-th expert with $L$-layer non-linear MLP, then we have:
\begin{align}
    f_s^{i} = F_{i}(h_{s}),~f_t^{i} = F_{i}(h_{t})
\end{align}
where $f_s^{i}\in \mathbb{R}^{d\times 1}$ and $f_t^{i}\in \mathbb{R}^{d\times 1}$ are the output of the $i$-th expert in source domain and target domain, respectively. Let $G_{s}$ and $G_{t}$ be the source domain gate and target domain gate that are non-linear MLP of $L$ layers with $\mathbb{R}^d \xrightarrow{} \mathbb{R}^d$, then we have:
\begin{align}
\label{eq:mmoe_gate}
    g_{s} = \text{softmax}(W_{s}^{gate}G_{s}(h_{s})),~g_{t} = \text{softmax}(W_{t}^{gate}G_{t}(h_{t}))
\end{align}
where $W_{s}^{gate}\in \mathbb{R}^{K\times d},W_{t}^{gate}\in \mathbb{R}^{K\times d}$ are the affine transformation and $K$ means the number of experts. $g_{s}\in \mathbb{R}^{K\times 1}$ and $g_{t}\in \mathbb{R}^{K\times 1}$ are the outputs of source domain gate and target domain gate.
With Eq.~\ref{eq:mmoe_gate}, the source and target CTR objective can automatically select its own parameters for optimization. To this end, we can obtain the latent features of two domains as $z_{s}\in \mathbb{R}^{d\times 1}$ and $z_{t}\in \mathbb{R}^{d\times 1}$:
\begin{align}
    z_{s}=\sum_{i=1}^{K} g_s^{i}f_s^{i},~z_{t}=\sum_{i=1}^{K} g_t^{i}f_t^{i}
\end{align}
With this MMOE-based feature extractor, on one hand, the parameter-sharing can help knowledge transfer. On the other hand, its architecture can effectively alleviate the optimization conflict issue~\cite{ma2018modeling}. This MMOE-based network is placed just behind the embedding layer because we believe the features are richer and generalized in low layers while more compact and specialized in high levels. The former one has more transferable patterns that can benefit downstream tasks.

\subsubsection{Cross-supervised Translator}
After obtaining the latent features of each domain, we propose a cross-supervised translator to learn the latent feature translation, which is inspired by image2image translation.
In image2image translation, images from target domain are translated as the images in the source domain, supervised by the true images in source domain. Whereas in CDCTR, the sample (\textit{e.g.} combined user feature and item feature) in the target domain may not share the user behavior label with a sample in the source domain, because these two samples do not have paired relationship.
Instead, a widely held belief in CDCTR is that if a user favors an item in the target domain, the favor behavior is preserved if the user and item are mapped into the source domain as corresponding features. For example, if a user likes science movies, he or she will also tend to love science novels.



In other words, given the latent feature $z_{t}$ of target domain, we aim to translate it into source domain as $z_{t}^{'}$ which can preserve the semantics of $z_{t}$. Then, $z_{t}^{'}$ can be taken into the prediction tower of source domain while supervised by the target domain label $y_{t}$.
Let $W_{t}^{tran}\in \mathbb{R}^{d\times d}$ be the translator of target domain and $BCE$ be the binary cross entropy loss, then the cross-supervised translator of target domain is optimized by:
\begin{align}
\label{eq:L_t_cross}
    \min~\mathcal{L}_{t}^{cross}=BCE(\sigma(R_{s}(z_{t}^{'})),y_{t}),~~z_{t}^{'} = W_{t}^{tran}z_{t}
\end{align}
where $\sigma$ is the \text{sigmoid} function and $R_{s}$ is the prediction tower of source domain. 
To stabilize the training, we have the symmetric formulation for source domain like Eq.~\ref{eq:L_t_cross} as:
\begin{align}
    \min~\mathcal{L}_{s}^{cross}=BCE(\sigma(R_{t}(z_{s}^{'})),y_{s}),~~z_{s}^{'} = W_{s}^{tran}z_{s}
\end{align}
where $R_{t}$ is the prediction tower of target domain and $W_{s}^{tran}\in \mathbb{R}^{d\times d}$ is the translator of source domain.
Note that the network architecture of $R_{s}$ and $R_{t}$ could be different architectures according to the domain's own characteristics.

In order to better conduct the latent feature translation,
apart from the above cross supervision, we add an orthogonal mapping constraint on the translators $W_{s}^{tran}$ and $W_{t}^{tran}$ as:
{\small
\begin{align}
\label{eq:orth_loss}
    \min~\mathcal{L}^{orth}=&||\mathbb{T}(W_{s}^{tran})(W_{s}^{tran}z_{s})-z_{s}||_{F}^{2}+ \nonumber \\
    &||\mathbb{T}(W_{t}^{tran})(W_{t}^{tran}z_{t})-z_{t}||_{F}^{2}
\end{align}
}%
where $\mathbb{T}$ is the transpose operation. The orthogonal transformation in mathematics has a characteristic that can preserve lengths and angles between vectors. Therefore, the orthogonal mapping constraint in Eq.~\ref{eq:orth_loss} can help the latent features $z_{t}$ of different samples preserve the similarity and avoid a case where multiple $z_{t}$ collapse to a single point after translation.

\subsubsection{Objective Function in Translation}
Apart from the above objective of learning translator, we still need the vanilla objective for optimizing the CTR task in each domain. Namely, the vanilla CTR objectives of source and target domain are formulated as:
{\small
\begin{align}
    \min \mathcal{L}_{s}^{vani}=BCE(\sigma(R_{s}(z_{s})),y_{s}),
    \mathcal{L}_{t}^{vani}=BCE(\sigma(R_{t}(z_{t})),y_{t})
\end{align}
}%
These two objectives help the model learn the network parameters (e.g. the tower network) targeted for CTR prediction. Meanwhile, when the tower networks are optimized for each domain's CTR prediction, the translators can adapt to the tower networks and learn feature translation in a meaningful and right direction. To sum up, the objective function of translation network is:
\begin{align}
\label{eq:translation_objective}
    \min \mathcal{L}_{trans}=\underbrace{\mathcal{L}_{s}^{vani}+\mathcal{L}_{t}^{vani}}_{\mathcal{L}^{vani}}+\alpha (\underbrace{\mathcal{L}_{s}^{cross}+\mathcal{L}_{t}^{cross}}_{\mathcal{L}^{cross}})+\beta \mathcal{L}^{orth}
\end{align}
where $\alpha$ and $\beta$ are hyper-parameters on loss weights. 

\subsection{Augmentation Network}
After the translation learning, we would first transfer the network parameters including the embedding layer, the shared MMOE module and the translator to the augmentation network. 
In order to give enough learning flexibility for augmentation network, as shown in Figure~\ref{figure:model_architecture}, the prediction tower is newly initialized rather than coming from the translation network.

\subsubsection{Cross-domain Augmentation} 
Meanwhile, the translator learned in translation network can map the latent feature $z_{t}$ into another space as $z_{t}^{'}$ and provide more useful information. That is to say, we can use the translated features of target domain samples as additional information to augment the target domain model training. The augmented latent feature of target domain is formulated as:
\begin{align}
    z_{t}^{aug}=z_{t}\oplus z_{t}^{'}
\end{align}
where $\oplus$ denotes the concatenation operation. When caring about the augmentation in source domain, the augmented feature can be obtained in a similar formula.

\subsubsection{Objective Function in Augmentation}
With the augmented latent feature $z_{t}^{aug}$, we would feed it into the new prediction tower $R_{t}^{aug}$ and use the vanilla CTR objective for fine-tuning. The objective is defined as:
\begin{align}
    \mathcal{L}_{aug}=BCE(\sigma(R_{t}^{aug}(z_{t}^{aug})),y_{t})
\end{align}
In this augmentation stage, the model has only one CTR objective and can avoid the optimization conflict problem in multi-objective models. When focusing on the performance of source domain, its augmentation network can be optimized in a similar way. The constraint in Eq.~\ref{eq:orth_loss} also works here with the same $\beta$ as in Eq.~\ref{eq:translation_objective}.

Considering the technique in knowledge transfer, the joint training works~\cite{hu2018conet,ma2018modeling,tang2020progressive,sheng2021one} and fine-tuning
methods~\cite{ouyang2020minet,liu2022continual,zhang2022keep} mainly rely on parameter-sharing, while our CDAnet proposes a novel translation then augmentation idea. In addition, CDAnet supports heterogeneous input features while most existing CDCTR models require homogeneous inputs.


\begin{table}[]
\centering
\caption{The statistics of datasets.}
\label{table:dataset}
\renewcommand{\arraystretch}{1.0}
 \setlength{\tabcolsep}{0.7mm}{ 
  \scalebox{0.85}{
\begin{tabular}{ccccc}
\hline
dataset               & \multicolumn{2}{c}{Amazon (movie and book)} & \multicolumn{2}{c}{Taobao (ad and rec)} \\ \hline
domain                & movie                & book                 & ad                 & rec                \\ \hline
\#users               & 29,680               & 52,690               & 141,917            & 186,731            \\
\#items               & 16,494               & 47,302               & 165,689            & 379,817            \\
\#input feature dim & 8,044                & 24,466               & 44,897             & 5,004              \\
\#train samples        & 1,685,836            & 7,133,107            & 3,576,414          & 12,168,878         \\
\#val samples              & 210,730              & 891,638              & 447,052            & 1,521,110            \\
\#test samples              & 210,730              & 891,639              & 447,052            & 1,511,110            \\
\#positive samples              & 351,216              & 1,486,064              & 234,736            & 855,362            \\ \hline
\end{tabular}
}}
\end{table}
\section{Experiments and Analysis}
\subsection{Experiment Setup}
\subsubsection{Datasets}
We first conduct our experiments on two public benchmarks whose two domains have heterogeneous input features: Amazon\footnote{https://jmcauley.ucsd.edu/data/amazon/} and Taobao\footnote{https://tianchi.aliyun.com/dataset/56}\footnote{https://tianchi.aliyun.com/dataset/649}. 
We choose the two largest domains--movie and book of Amazon to conduct the experiments. In the movie domain, we have the user ID, movie ID, movie genre, movie director and movie name information. While in the book domain, we have the user ID, book ID, book category, book writer and book name information, which are varying from those in the movie domain. Also, each domain has its own user behaviors. The original user behaviors are 0-5 ratings and we process the scores larger than 3 as positive feedback and other scores as negative feedback for CTR prediction. The user and item ID are both embedded as 64-dim features. Both movie and book name are processed as vectors by a word2vec Glove-6B model\footnote{https://nlp.stanford.edu/projects/glove/}. The movie genre, movie director, book category and book writer are mapped as one-hot features, respectively for each domain. 
Taobao dataset contains the user-item interactions of the advertisement (ad) and recommendation (rec) domain. We consider the buy behavior type as positive feedback and other types as negative feedback. In the ad domain, we get the user ID, age, gender, occupation and some other user profile information for the user side. As for the ad side, we have the ad ID, ad category and ad brand information. In the rec domain, we get user ID, item ID and item category information, which shows quite different features from the ad domain. For Taobao dataset, the user and item ID are both embedded as 64-dim features. Other discrete features are processed as one-hot or multi-hot features. 
For both datasets, we also apply a $k$-core filtering to guarantee each user or item has at least $k$ interactions. $k$ is 5 and 10 for movie and book domain of Amazon, respectively. For Taobao, $k$ also equals 5 and 10 for ad and rec domain, respectively. The dataset statistics are summarized in Table~\ref{table:dataset}.

Furthermore, since we address the real-world CDCTR with heterogeneous input features at Taobao, we also evaluate our model on Alibaba production data and online experiments on Taobao mobile app. In particular, we collect six-month user behaviors in text2product retrieval as the source domain data and one-year user behaviors in image2product retrieval as the target domain data. Both domains contain hundreds of billions of samples and rich user-side and item-side input features that are used in the online production system. The input features in these two domains are quite different due to the characteristic of these scenarios.

\subsubsection{Baselines}
Existing pre-training \& fine-tuning methods (e.g. Keep~\cite{zhang2022keep}, CTNet~\cite{liu2022continual}) require shared feature fields for knowledge transfer, so we do not include them as the baselines. Instead, we adopt various methods for comparison, including single domain and joint training methods: 
\begin{itemize}
    \item MLP. A deep multi-layer perception model is a common and efficient ranking model in online search and recommendation systems. Here, MLP serves as a single-domain CTR model for each domain.
    \item ShareMiddle. Considering the input features across domains are heterogeneous, we separate the embedding layers while sharing the middle layer just after the embedding layer of each domain inspired by the ShareBottom~\cite{ma2018modeling} technique.
    \item STAR~\cite{sheng2021one}. STAR is a star topology model that trains a single model to serve all domains by leveraging the data from all domains simultaneously.
    \item DDTCDR~\cite{li2019ddtcdr}. It is a deep dual transfer learning model that transfers knowledge between related domains in an iterative manner.
    \item MMOE~\cite{ma2018modeling}. MMOE implicitly models task relationships for multi-task learning by a shared mixture-of-experts module and task-specific gates. We utilize MMOE here for  CDCTR with domain-specific embedding layers.
    \item PLE~\cite{tang2020progressive}. PLE is a multi-task learning model that separates shared components and task-specific components explicitly and adopts a progressive routing mechanism to extract and transfer knowledge.
\end{itemize}

\subsubsection{Parameter Settings}
In our experiments on public benchmarks, we split the data into train, validation, and test sets with the common 8:1:1 setting according to chronological order. The experiments are conducted multiple times and the mean value is taken as the model performance. We set the latent feature size as 64 for all models. The number of training epochs is set as 200 which can ensure the model's convergence. To make a fair comparison, we conduct a grid search of hyper-parameters and the number of layers for all models. Considering CDAnet, for both domains of Amazon dataset, $\alpha$ is 0.01, $\beta$ equals 0.1, the number of experts is 2 and each expert has 2 layers, the prediction tower has 2 layers including the logit mapping layer. 
For both domains of Taobao dataset, $\alpha$ is 0.03, $\beta$ equals 0.1, the number of experts is 2, each expert has 2 layers and the prediction tower has 3 layers including the logit mapping layer. On our production dataset, the prediction tower of text2product domain is a DCN-v2~\cite{wang2021dcn} model\footnote{We have no access to the online production model of text2product domain, so we empirically choose DCN-v2 as the prediction tower network.} while that of our image2product domain is an online serving model. Based on the experience on public benchmarks, we also set the number of experts is 2, each expert has 2 layers, $\alpha$ and $\beta$ are both 0.01 for our production dataset\footnote{We do not tune the hyper-parameters on this production dataset since it would cost too much computation resource in a limited time. Different hyper-parameters would be tuned in future experiments.}.


\begin{table}[]
\centering
\caption{The AUC comparison of different models on Amazon and Taobao. The results of both domains are listed.}
\label{table:overall_public}
\begin{tabular}{ccccc}
\hline
Dataset     & \multicolumn{2}{c}{Amazon} & \multicolumn{2}{c}{Taobao}                                                      \\ \hline
Domain      & Movie            & Book    & Ad                                     & Rec                                    \\ \hline
MLP         & 0.6595           & 0.7604  & 0.6161                                 & 0.6865                                 \\
ShareMiddle & 0.6604           & 0.7603  & 0.6160                                  & 0.5020                                  \\
STAR        & 0.6503           & 0.7626  & 0.6149                                 & 0.6795                                 \\
DDTCDR      & 0.6636           & 0.7807  & 0.6162                                 & 0.6756                                 \\
MMOE        & 0.7025           & \textbf{0.7899}  & 0.6177                                 & 0.6930                                 \\
PLE         & 0.6993           & 0.7846  & 0.6164                                 & 0.6933                                 \\
CDAnet       & \textbf{0.7225}  &  0.7811       & \textbf{0.6200} & \textbf{0.7034} \\ \hline
\end{tabular}
\end{table}


\begin{table}[]
\centering
\caption{The AUC comparison results of image2product domain on our production dataset. Here we consider text2product as the source domain and our image2product domain as the target domain.}
\label{table:overall_industrial}
\begin{tabular}{cc}
\hline
Task  & CTR        \\ \hline
Base  & 0.7845  \\
CDAnet & 0.7866(\textbf{+0.21}\%)           \\ \hline
\end{tabular}
\end{table}

\subsection{Overall Comparison}
\subsubsection{Offline Comparison}
In this part, we show the comparison results on both domains of the datasets. By following popular works~\cite{liu2022continual,sheng2021one}, we take AUC as the evaluation metric. The results are shown in Table~\ref{table:overall_public} and Table~\ref{table:overall_industrial}. 

From Table~\ref{table:overall_public}, we observe that CDAnet generally can improve the CTR performance compared to various models. The joint training models (e.g. DDTCDR, MMOE, PLE) originally are not supportive for CDCTR with heterogeneous input features, while we revise them with domain-specific embedding layers so that they can work. Their inferior performance to CDAnet indicates that simply revising existing models with domain-specific embedding layers cannot well transfer the knowledge across domains. Although STAR can work with heterogeneous inputs, it mainly relies on a centered and domain-shared network to transfer knowledge, which is not so effective according to our results in Table~\ref{table:overall_public}. Instead, our translation then augmentation idea is superior and can provide better performance.
In addition, considering the results in Table~\ref{table:overall_industrial}, we see CDAnet benefits CTR task even with so extremely large-scale data. It brings an absolute 0.21\% AUC increase for CTR. A possible reason for this is that CDAnet can well transfer the knowledge of large-scale and high-quality data in text2product domain to image2product domain. 

\begin{figure*}[t]
\centering
\begin{minipage}[t]{0.44\textwidth}
\centering
\includegraphics[width=\textwidth]{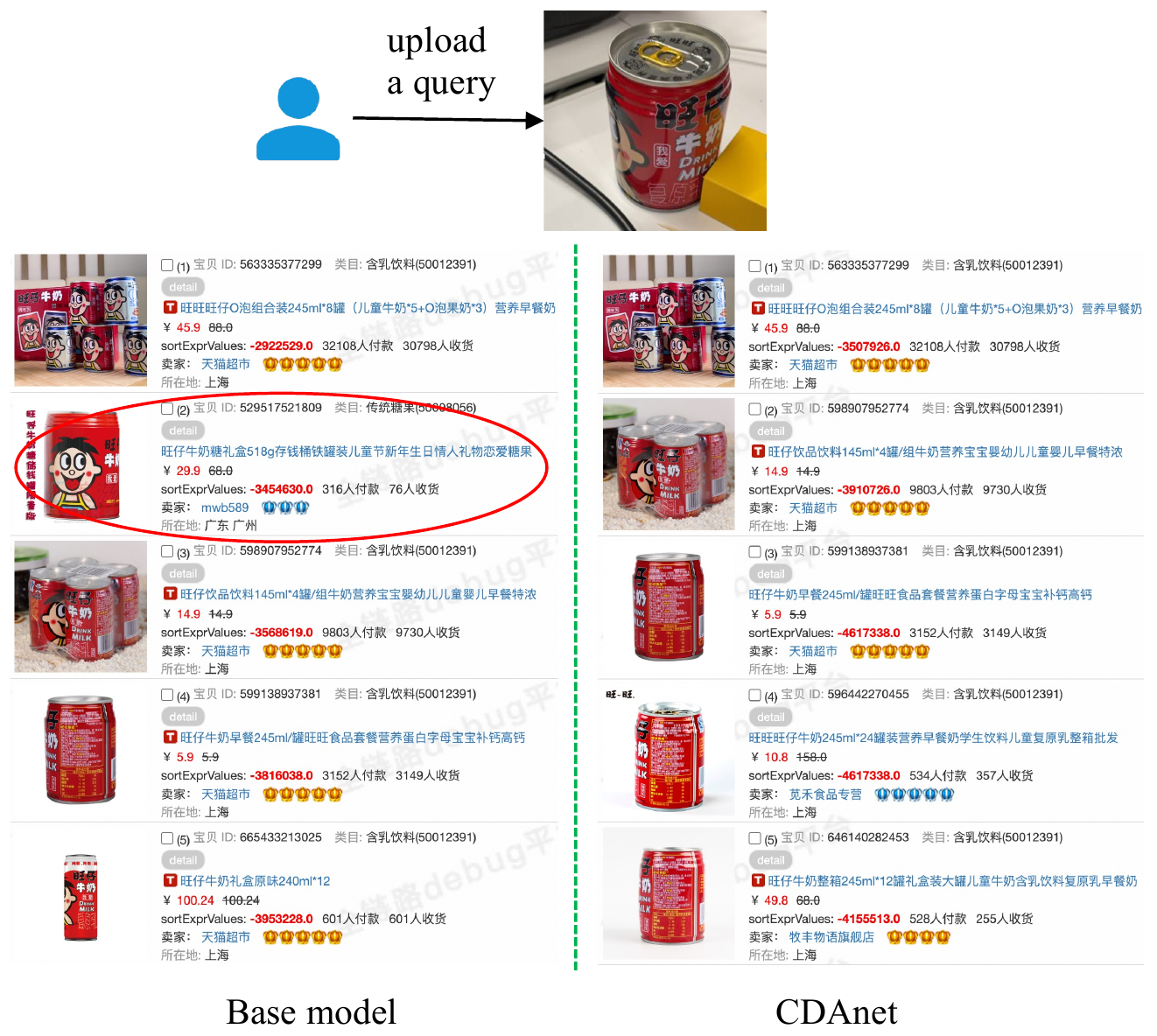}
\caption*{(a) \footnotesize{example of Wangzai milk}}
\end{minipage}
\hspace{10pt}
\begin{minipage}[t]{0.44\textwidth}
\centering
\includegraphics[width=\textwidth]{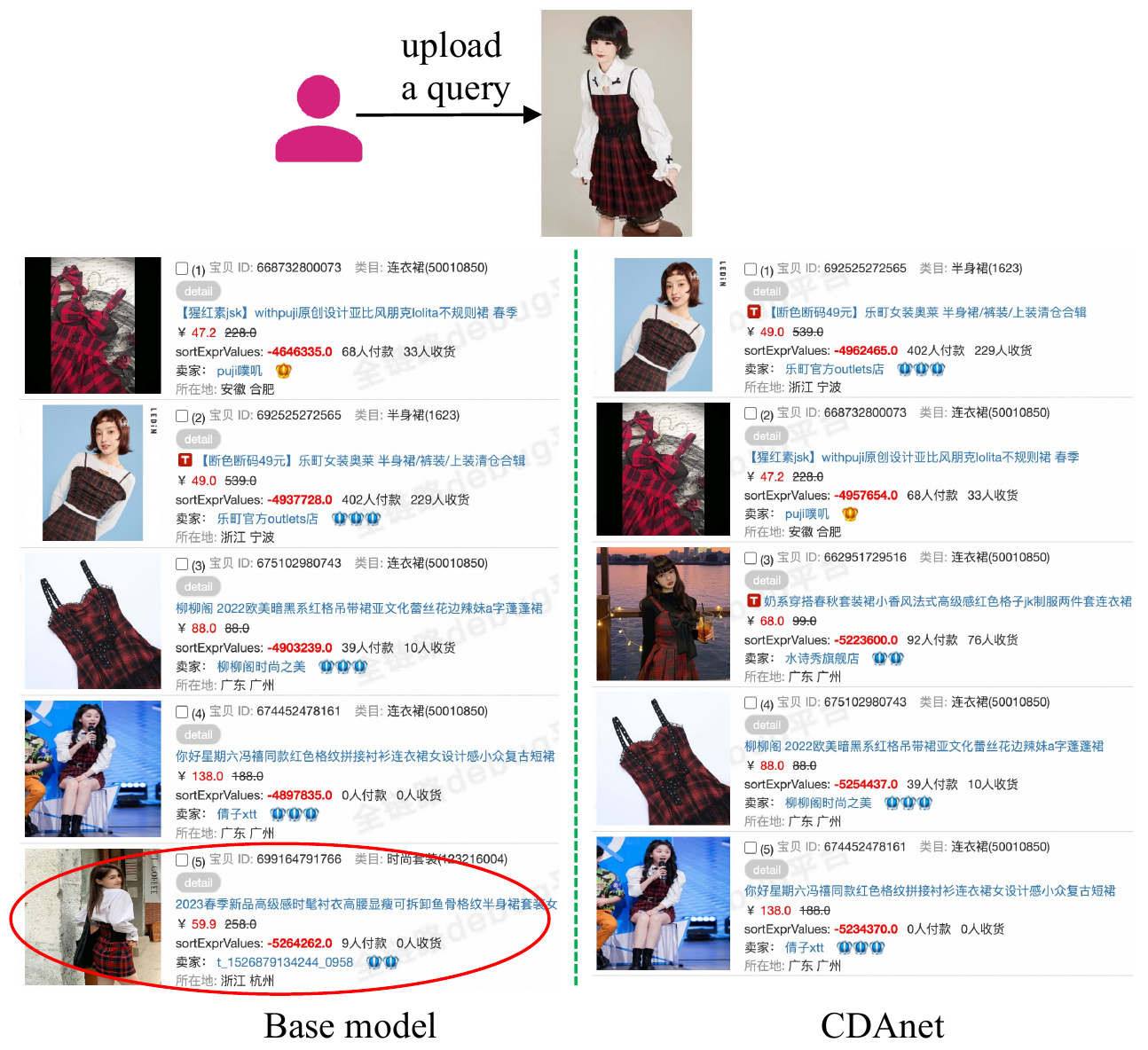}
\caption*{(b) \footnotesize{example of a dress}}
\end{minipage}
\caption{The ranking examples of the online base model and our proposed CDAnet. Given a query with the same user, we list the top5 ranking items of different models. (a) is a example of Wangzai milk and (b) is an example of a dress.}
\label{figure:ranking_exmples}
\end{figure*}
\subsubsection{Online A/B test}
We further conduct online experiments in an A/B test framework of image2product retrieval at Taobao app over 20 days. The baseline model is the online serving model trained on only image2product data. The online evaluation metrics are real CTR and GMV. The online A/B test shows that our CDAnet leads to an absolute 0.11 point CTR increase and a relative 1.26\% GMV increase. In addition, we compare some online ranking examples between the baseline model and our CDAnet. The result is shown in Figure~\ref{figure:ranking_exmples}. 
In this figure, we give two examples to illustrate the better ranking ability of our CDAnet. In (a), if the user uploads a picture of Wangzai milk, we see that the online base model ranks a bottle of Wangzai candy that is quite visually similar in position 2. Instead, our CDAnet can exclude this case and the top5 items are all about Wangzai milk. A possible reason may be that the query of milk and candy in text domain are quite different. The CTR model in text2product retrieval can well capture this while the CTR model in image2product domain may easily be confused by the visual patterns. Our CDAnet can transfer the knowledge of text2product domain into our image2product domain and improve the ranking results. In (b), we show a ranking example of a dress. Although our database does not have the the same item as the query, CDAnet has better ability of ranking similar items as the dress, while the online base model has a skirt in position 5. These examples can help us better understand the superiority of CDAnet.


\begin{figure*}[t]
\centering
\begin{minipage}[t]{0.24\textwidth}
\centering
\includegraphics[width=\textwidth]{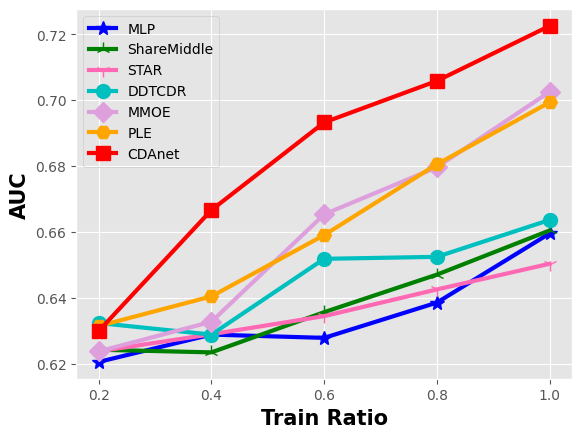}
\caption*{(a) \footnotesize{Amazon-movie}}
\end{minipage}
\begin{minipage}[t]{0.24\textwidth}
\centering
\includegraphics[width=\textwidth]{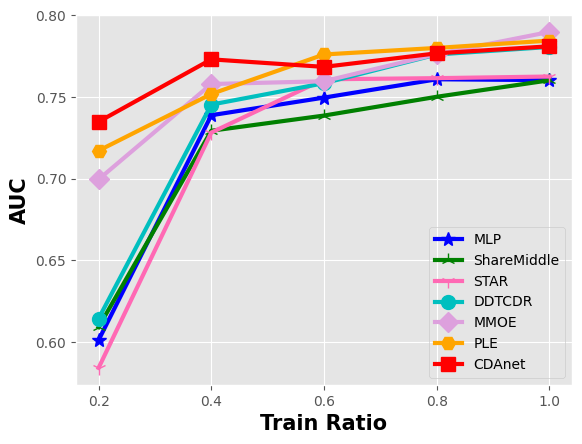}
\caption*{(b) \footnotesize{Amazon-book}}
\end{minipage}
\begin{minipage}[t]{0.24\textwidth}
\centering
\includegraphics[width=\textwidth]{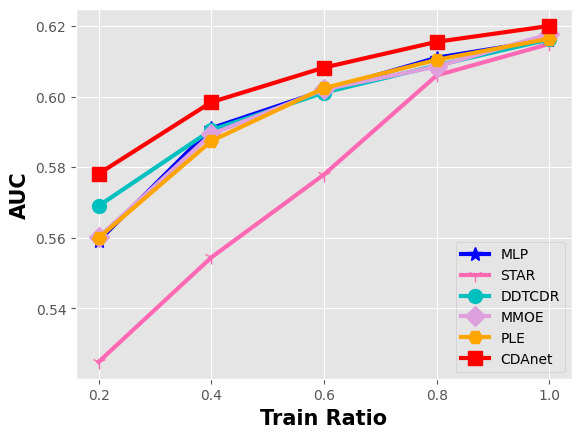}
\caption*{(c) \footnotesize{Taobao-ad}}
\end{minipage}
\begin{minipage}[t]{0.24\textwidth}
\centering
\includegraphics[width=\textwidth]{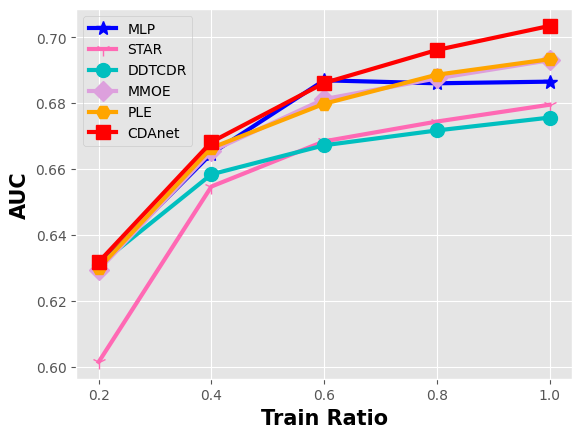}
\caption*{(d) \footnotesize{Taobao-rec}}
\end{minipage}
\caption{The effects of different sparsity levels on Amazon and Taobao. Train ratio means the ratio of the original train data.}
\label{figure:sparse_data}
\end{figure*}
\subsubsection{Different Sparse Data}
In order to investigate how CDAnet performs under more sparse conditions, we conduct an experiment to see the model performance at different sparsity levels of the training data.
In particular, fixing the validation and test set, we vary the ratio of the original train data to the new train set. 

The results are shown in Figure~\ref{figure:sparse_data}, where we can see that CDAnet can achieve better results at almost all sparsity levels compared to other cross-domain CTR methods. CDAnet has a more advanced knowledge transfer style that considers both the heterogeneous inputs and optimization conflict problem, so that is can show more generalized performance.

\begin{table*}[]
\centering
\caption{Results to show that the translated features in book space have related item semantics as its original item semantics in movie space. The title is the movie name or book name. Bolded book titles mean close semantics between the movie and book.}
\label{table:translated_analysis}
\renewcommand{\arraystretch}{1.2}
 \setlength{\tabcolsep}{0.7mm}{ 
  \scalebox{0.85}{
\begin{tabular}{c|cc|cc|c}
\hline
                       & \multicolumn{2}{c|}{items in movie domain}                                                                  & \multicolumn{2}{c|}{items of 5-nearest neighbours from Book domain}                                                      &                                     \\ \hline
UserID                 & \multicolumn{1}{c|}{ItemID}                 & Title                                         & \multicolumn{1}{c|}{ItemId} & Title                                            & item genre                    \\ \hline
\multirow{5}{*}{21778} & \multicolumn{1}{c|}{\multirow{5}{*}{8329}}  & \multirow{5}{*}{The Addams Family}            & \multicolumn{1}{c|}{39040}  & \textbf{Where My Heart Breaks}                   & \multirow{5}{*}{emotion, love}      \\ \cline{4-5}
                       & \multicolumn{1}{c|}{}                       &                                               & \multicolumn{1}{c|}{33024}  & The Program                                      &                                     \\ \cline{4-5}
                       & \multicolumn{1}{c|}{}                       &                                               & \multicolumn{1}{c|}{29368}  & \textbf{Lick (A Stage Dive Novel)}               &                                     \\ \cline{4-5}
                       & \multicolumn{1}{c|}{}                       &                                               & \multicolumn{1}{c|}{37926}  & \textbf{Snowed Over}                             &                                     \\ \cline{4-5}
                       & \multicolumn{1}{c|}{}                       &                                               & \multicolumn{1}{c|}{44092}  & \textbf{Where the Stars Still Shine}             &                                     \\ \hline
\multirow{5}{*}{4842}  & \multicolumn{1}{c|}{\multirow{5}{*}{13887}} & \multirow{5}{*}{Mystery Science Theater 3000} & \multicolumn{1}{c|}{10505}  & \textbf{A Wild Sheep Chase: A Novel}             & \multirow{5}{*}{mystery, thrillers} \\ \cline{4-5}
                       & \multicolumn{1}{c|}{}                       &                                               & \multicolumn{1}{c|}{7713}   & \textbf{Wyrd Sisters}                            &                                     \\ \cline{4-5}
                       & \multicolumn{1}{c|}{}                       &                                               & \multicolumn{1}{c|}{8205}   & \textbf{Polar Star}                              &                                     \\ \cline{4-5}
                       & \multicolumn{1}{c|}{}                       &                                               & \multicolumn{1}{c|}{135765} & \textbf{Harry Potter and the Chamber of Secrets} &                                     \\
                       & \multicolumn{1}{c|}{}                       &                                               & \multicolumn{1}{c|}{6374}   & \textbf{The Ambler Warning}                      &                                     \\ \hline
\multirow{5}{*}{2078}  & \multicolumn{1}{c|}{\multirow{5}{*}{4977}}  & \multirow{5}{*}{Stargate SG-1 Season 2}       & \multicolumn{1}{c|}{43606}  & Countess So Shameless                            & \multirow{5}{*}{fantasy, romance}   \\ \cline{4-5}
                       & \multicolumn{1}{c|}{}                       &                                               & \multicolumn{1}{c|}{16339}  & \textbf{Crimson City}                            &                                     \\ \cline{4-5}
                       & \multicolumn{1}{c|}{}                       &                                               & \multicolumn{1}{c|}{35940}  & \textbf{Not Your Ordinary Wolf Girl}             &                                     \\ \cline{4-5}
                       & \multicolumn{1}{c|}{}                       &                                               & \multicolumn{1}{c|}{10612}  & \textbf{The Resisters}                           &                                     \\ \cline{4-5}
                       & \multicolumn{1}{c|}{}                       &                                               & \multicolumn{1}{c|}{12671}  & \textbf{Dagger-Star (Epic of Palins, Book 1)}    &                                     \\ \hline
\end{tabular}
}}
\end{table*}
\subsection{The Semantics of Translated Features}
CDAnet learns how to translate the latent features between domains and exploits these translated latent features as additional information to boost the target domain CTR prediction, so it is curious to see whether the translated features are meaningful for target domain CTR prediction. In this part, we design an experiment to see whether the latent features have related semantics before and after translation. 
To be specific on Amazon dataset, given a user that has training instances sharing positive labels in both domains
\footnote{Notice our method does not require a user to have training instances in both domains. We make this requirement here just to better analyze the translated features. We let the instances in two domains share the positive label here for analysis because the ``negative'' instances in CTR are usually sampled and have less confidence than positives.}, we can obtain the latent feature matrix $Z_{b}\in \mathbb{R}^{N_{b}\times d}$ of book domain and $Z_{m}\in \mathbb{R}^{N_{m}\times d}$ of movie domain.
Then, we can get the translated features of $Z_{m}$ by the learned translator and it is denoted as $Z_{m}^{'}\in \mathbb{R}^{N_{m}\times d}$. 
Next, for a row in the the translated feature $Z_{m}^{'}$, we find its $k$-nearest neighbours in $Z_{b}$. In this case, we denote the book names of these neighbors as the corresponding information of the translated feature. Finally, for a row in $Z_{m}^{'}$, we can check whether the book name of the neighbors in book space have related semantics with the original movie name in movie space. The results are summarized in Table~\ref{table:translated_analysis}.

From this Table, we find that the items of 5-nearest neighbours from book domain have close semantics to the corresponding item in the movie domain. For example, given UserId ``21778'', the interacted movie is ``8329'' and the movie name is ``The Addams Family'' which tells an emotional and love story. Meanwhile, for the corresponding translated features, the items of 5-nearest neighbors from book domain are novels also about emotion and love. It shows that the translated features can capture related semantics from movie domain to book domain. This result matches a common belief in CDCTR that if a user likes an item in one domain, the user may also love the items sharing related semantics in other domains~\cite{li2019ddtcdr,liu2022continual,chen2020towards}. Notice that our model does not use any alignment information of items across domains, but it can automatically capture this and provide more useful information to boost the CTR prediction performance. This result further demonstrates the value of our translation network.

\begin{table}[]
\centering
\caption{The ablation study on different model parts of CDAnet. The figures are AUC results. ``w/o'' means without the corresponding model part. ``w/o translation network'' means we directly train the augmentation network which is random initialized. ``w/o augmentation network'' indicates we just use the trained translation network for evaluation.}
\label{table:different_modules}
\begin{tabular}{ccccc}
\hline
Dataset                   & \multicolumn{2}{c}{Amazon} & \multicolumn{2}{c}{Taobao} \\ \hline
Domain                    & movie        & book        & ad           & rec         \\ \hline
w/o MMOE                  & 0.7186       & 0.7648      & 0.6199       & 0.7012      \\
w/o $\mathcal{L}^{orth}$ & 0.7153       & 0.7734      & 0.6190      & 0.7022      \\
w/o $\mathcal{L}^{cross}$ & 0.7188       & 0.7757      & 0.6191      & 0.7003      \\
w/o translation network     & 0.6464       & 0.7649      & 0.6075            & 0.7001           \\
w/o augmentation network    & 0.6818       & 0.7686      & 0.6168       & 0.6811      \\
CDAnet                     & \textbf{0.7225}       & \textbf{0.7811}      & \textbf{0.6200}      & \textbf{0.7034}      \\ \hline
\end{tabular}
\end{table}
\subsection{Ablation Study}
\subsubsection{The Impacts of Different Model Parts}
In order to assess the consequences of varying model components, we undertook an experiment to analyze the effect on performance upon the removal of the associated module. The results are shown in Table~\ref{table:different_modules}.

Comparing the result between CDAnet and w/o MMOE, we see the MMOE-based feature extractor has positive effects on boosting model performance. This MMOE module 
may help translation network alleviate optimization conflict issue and better transfer knowledge. Considering the result of w/o $\mathcal{L}^{orth}$ and w/o $\mathcal{L}^{cross}$,
we see either removing the orthogonal constraint loss $\mathcal{L}^{orth}$ or removing the cross-supervision loss $\mathcal{L}^{cross}$ would cause deteriorated performance. The orthogonal constraint loss $\mathcal{L}^{orth}$ can help CDAnet keep the similarity among latent features after translation and avoid mode collapse problem. The cross-supervision loss $\mathcal{L}^{cross}$ plays a vital role in learning the two translators $W_{s}^{trans}$ and $W_{t}^{trans}$. Without $\mathcal{L}^{cross}$, the translators cannot learn how to translate latent features into another space. 

Further, either without translation network or without augmentation network would cause rather poor performance. 
The translation network learns how to transfer knowledge between domains. 
When removing it, we cannot have useful knowledge for later augmentation network. 
The augmentation network reuses the pre-trained parameters of translation network and employs the additional translated latent features for final CTR prediction goal. When removing the augmentation network, we only have the translation network whose goal is translation and cannot guarantee good CTR prediction performance.

\begin{figure}[t]
\centering
\begin{minipage}[t]{0.2\textwidth}
\centering
\includegraphics[width=\textwidth]{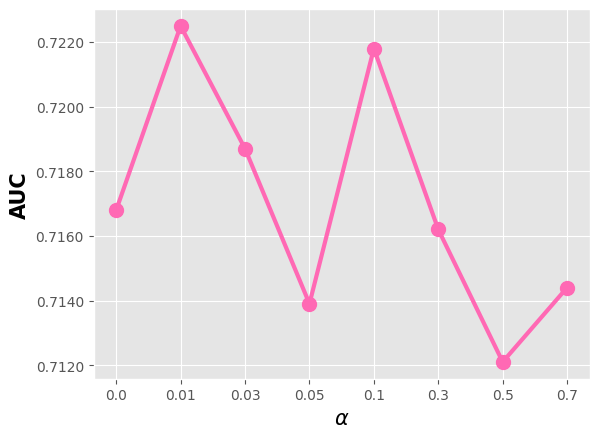}
\caption*{(a) \footnotesize{$\alpha$ on Amazon-movie}}
\end{minipage}
\begin{minipage}[t]{0.2\textwidth}
\centering
\includegraphics[width=\textwidth]{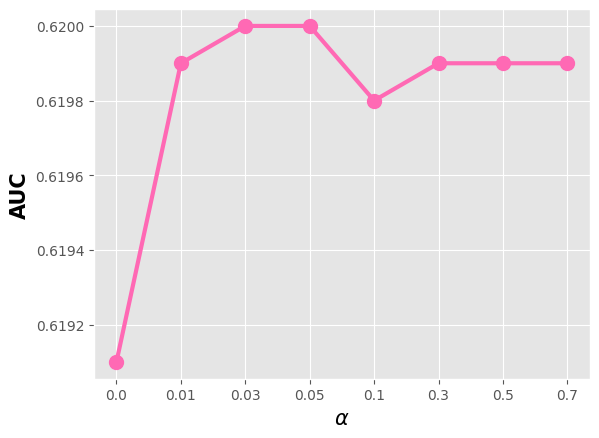}
\caption*{(b) \footnotesize{$\alpha$ on Taobao-ad}}
\end{minipage}\\
\begin{minipage}[t]{0.2\textwidth}
\centering
\includegraphics[width=\textwidth]{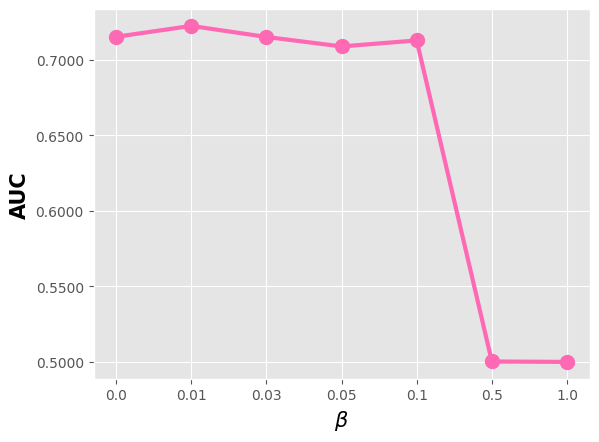}
\caption*{(c) \footnotesize{$\beta$ on Amazon-movie}}
\end{minipage}
\begin{minipage}[t]{0.2\textwidth}
\centering
\includegraphics[width=\textwidth]{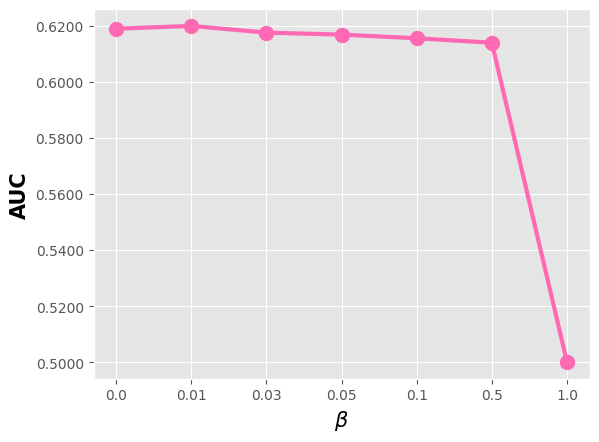}
\caption*{(d) \footnotesize{$\beta$ on Taobao-ad}}
\end{minipage}\\
\begin{minipage}[t]{0.2\textwidth}
\centering
\includegraphics[width=\textwidth]{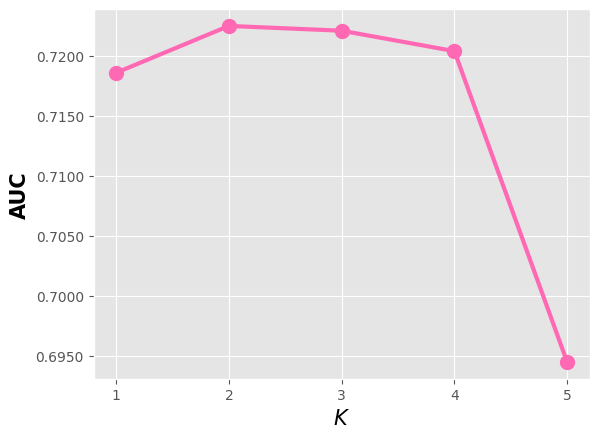}
\caption*{(e) \footnotesize{$K$ on Amazon-movie}}
\end{minipage}
\begin{minipage}[t]{0.2\textwidth}
\centering
\includegraphics[width=\textwidth]{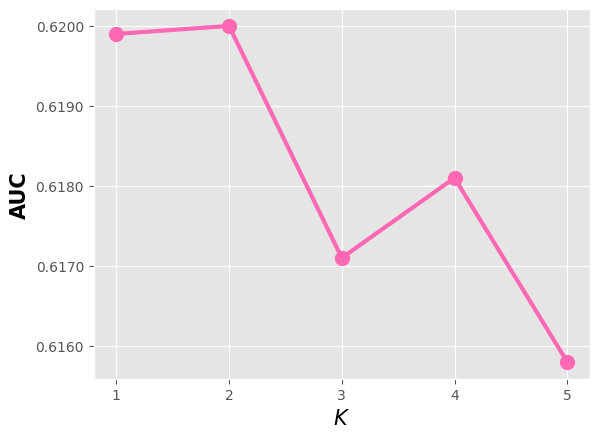}
\caption*{(f) \footnotesize{$K$ on Taobao-ad}}
\end{minipage}
\caption{The effects of hyper-parameters on different datasets. $\alpha$ and $\beta$ are loss weights on different loss parts. $K$ denotes the number of experts.}
\label{figure:different_hyperparameters}
\end{figure}
\subsubsection{Parameter Sensitivity}
In CDAnet, $\alpha$ and $\beta$ control the loss weight on cross-supervision loss and orthogonal constraint loss, respectively. $K$ controls the number of experts in the shared MMOE-based feature extractor. Here, we study the parameter sensitivity of these hyper-parameters to investigate their effects on model performance. The results are shown in Figure~\ref{figure:different_hyperparameters}. 

In this figure, $\alpha$ controls the strength of cross-supervision on learning the translators. Too large $\alpha$ may dominate the learning of $\mathcal{L}^{vani}$ and cause bad effects on learning cross-supervised translators. $\beta$ controls the weight of $\mathcal{L}^{orth}$ and too large $\beta$ may limit the learning flexibility of the translators. $K$ is the number of experts in the shared MMOE-based feature extractor and $K=2$ shows the best performance. A possible reason may be too large $K$ involves too many parameters and the data does not support the model's complexity.

\section{Conclusion and Future Work}
Cross-domain click-through rate (CDCTR) prediction with heterogeneous input features is an important and practical problem in real-world systems. In this paper, we propose a novel model named CDAnet that contains a translation network and augmentation network for effective knowledge transfer in CDCTR with heterogeneous input features. Through extensive experiments, we show CDAnet is able to learn meaningful translated latent features and boost the CTR prediction performance. Results on the large-scale production dataset and online system at Taobao app show its superiority in real-world applications. 

Although CDAnet has achieved better performance than existing models, it still has inadequacies in transferring knowledge. The latent features of two domains are usually not fully overlapped, which means some ingredients of the latent features may not be translated and may have negative effects in translation. In the future, we will study a more effective technique for latent feature translation and boost the target domain CTR prediction.


\bibliographystyle{IEEEtran}
\bibliography{IEEEabrv,paper}

\end{document}